\documentclass[onecolumn,usenatbib,useAMS]{mn2e}

\usepackage{natbib}
\usepackage{graphicx,epsfig}
\usepackage{latexsym,amssymb}
\usepackage[fleqn]{amsmath}

\newcommand{\aj}{AJ}% 
\newcommand{\apj}{ApJ}% 
\newcommand{\apjs}{ApJS}% 
\newcommand{\aap}{A\&A}% 
\newcommand{\mnras}{MNRAS}% 
\newcommand{\nat}{Nat}% 
\newcommand{\pasj}{PASJ}% 
\newcommand\avg[1]{\left\langle{#1}\right\rangle}
\newcommand\order[2]{{\cal O}\left({#1}^{#2}\right)}

\newcommand\img{{\rm img}}
\newcommand\Rein{R_{\rm ein}}
\newcommand\Sigcr{\Sigma_{\rm crit}}

\newcommand\vx{\mbox{\boldmath $x$}}
\newcommand\vu{\mbox{\boldmath $u$}}
\newcommand\vv{\mbox{\boldmath $v$}}
\newcommand\va{\mbox{\boldmath $\alpha$}}

\newcommand\Cmat{{\bf C}}
\newcommand\Gmat{{\bf \Gamma}}
\newcommand\Imat{{\bf I}}
\newcommand\Mmat{{\bf M}}

\newcommand\cov{\mbox{cov}}
\newcommand\tr{{\rm tr}}

\newcommand\ksub{\bar\kappa_s}
\newcommand\khat{\hat\kappa_s}
\newcommand\kimg{\bar\kappa_{s,\img}}
\newcommand\rimg{r_\img}
\newcommand\timg{\theta_\img}

\newcommand\rhat{\hat r}
\newcommand\ghat{\tilde\gamma}
\newcommand\ahat{\tilde\alpha}
\newcommand\mbar{\avg{m}}
\newcommand\meff{m_{\rm eff}}
\newcommand\dmu{\delta{\mskip-2mu}\mu}
\newcommand\smth{^{0}}
\newcommand\sub{^{s}}

\newcommand\refeq[1]{eq.~(\ref{eq:#1})}
\newcommand\refeqs[2]{eqs.~(\ref{eq:#1}) and (\ref{eq:#2})}
\newcommand\reffig[1]{Figure~\ref{fig:#1}}
\newcommand\reftab[1]{Table~\ref{tab:#1}}
\newcommand\refsec[1]{\S~\ref{sec:#1}}

\newcommand\refapp[1]{Appendix~\ref{app:#1}}

\title[Lensing with stochastic substructure]
{Gravitational lensing with stochastic substructure: Effects of the clump mass function and spatial distribution}

\author[Keeton]{
Charles R.\ Keeton \\
Department of Physics and Astronomy, Rutgers University, 136 Frelinghuysen Road, Piscataway, NJ 08854 USA
}

\begin{document}

%\date{\today}

\maketitle 

\begin{abstract}
Mass clumps in gravitational lens galaxies can perturb lensed images in characteristic ways.  Strong lens flux ratios have been used to constrain the amount of dark matter substructure in lens galaxies, and various other observables have been considered as additional probes of substructure.  We study the general theory of lensing with stochastic substructure in order to understand how lensing observables depend on the mass function and spatial distribution of clumps.  We find that magnification perturbations are mainly sensitive to the total mass in substructure projected near the lensed images; when the source is small, flux ratios are not very sensitive to the shape of the clump mass function.  Position perturbations are mainly sensitive to a characteristic clump mass scale, namely $\meff = \avg{m^2}/\avg{m}$, with some mild dependence on other mass moments when the spatial distribution is not uniform.  They have contributions from both ``local'' and ``global'' populations of clumps (i.e., those projected near the images, and those farther away).  Time delay perturbations are sensitive to the same characteristic mass, $\meff$, and mainly driven by the global population of clumps.  While there is significant scatter in all lensing quantities, there are some non-trivial correlations that may contain further information about the clump population.  Our results indicate that a joint analysis of multiple lens observables will offer qualitatively new constraints on the mass function and spatial distribution of dark matter substructure in distant galaxies.
\end{abstract}

%\begin{keywords}
%\end{keywords}

%=====================================================================
\section{Introduction}
\label{S:intro}
%=====================================================================

The flux ratios in 4-image gravitational lens systems have been used to constrain the amount of dark matter substructure in lens galaxies \citep[e.g.,][]{MS98,MM01,chiba02,DK02}.  Recently there has been considerable interest in identifying other types of lensing observations that can provide additional information about the population of dark matter clumps.  For example, with compact sources (i.e., quasars) the possibilities include:
\begin{itemize}
\item Flux ratios measured at multiple wavelengths corresponding to different source sizes \citep[e.g.,][]{MoustakasMetcalf,metcalf2237,chiba-IR,dobler-finsrc,macleod-IR,minezaki-IR}.
\item Precise image positions \citep[e.g.,][]{koopmans-2016,chen-pos,bisector,more-2016}.
\item High-resolution radio interferometry to resolve the images into multiple ``milli-images'' \citep[e.g.,][]{gorenstein-0957,garrett-0957,trotter-0414,ros-0414,biggs-0128} and search for possible image splittings induced by substructure \citep[e.g.,][]{yonehara,riehm1,riehm2}.
\item Precise time delays between the images \citep[e.g.,][]{morgan-1131,NewChannel,congdon-tdel}.
\end{itemize}
With extended sources it is possible to look for distortions in the shapes and/or surface brightness distributions of the images \citep[e.g.,][]{metcalf-bent,inoue1,inoue2,gravimg1,gravimg2,gravimg3}.

The high-level goal of such diverse observations is to measure not only the mean density of dark matter substructure but also the mass function, spatial distribution, and perhaps redshift evolution of the clump population.  Those are the key quantities for testing predictions from the Cold Dark Matter paradigm \citep[recent examples include][]{VL-sub,VLII,aquarius} and placing astrophysical constraints on the particle nature of dark matter \citep[e.g.,][]{moustakas-decadal}.

The time is ripe to develop a comprehensive theoretical framework for substructure lensing that consolidates the various observables and reveals how they depend on physical properties of the clump population.  The relevant framework is that of stochastic lensing, in which we treat the positions and masses of the clumps as random\footnote{The dark matter substructure in a galaxy is not truly random; in principle it is determined by the galaxy's formation history.  However, the formation process is sufficiently complicated, and impossible to reconstruct, that it is fair to treat the substructure as random.} variables and compute statistical properties of lensing observables.  While the application to dark matter substructure is new, there has been considerable formal work on stochastic lensing in the context of stellar microlensing.  \citet{deguchi1,deguchi2} computed the variance of brightness fluctuations.  \citet{nityananda} and \citet{schneider-shear} derived the probability distribution for the lensing shear, which \citet{schneider-mu} used to find the probability distribution for the magnification.  Several studies considered the probability distribution for lensing deflections especially as they relate to the statistics of  microlensing light curves \citep{katz,seitz1,seitz2,neindorf,tuntsov1,tuntsov2}.  \citet{alberto1,alberto2} recently began a rigorous program to derive probability distributions for many different lensing quantities for an arbitrary number of stars (i.e., not necessarily in the limit of $N \to \infty$).

All of those studies assumed the mass clumps have a probability distribution that is spatially uniform.  While that approximation is reasonable on the scales that are relevant for stellar microlensing, it may not be appropriate for applications related to dark matter substructure.  Our goal in this paper is to generalise aspects of the theory of stochastic lensing to allow arbitrary spatial distributions of mass clumps.  We also consider arbitrary mass functions, which have been explicitly examined in only some of the microlensing work \citep[e.g.,][]{katz,neindorf}.

Regardless of the spatial distribution and mass function, the lensing effects from a collection of mass clumps can be written as a superposition of effects from individual clumps.  Since the resulting sums contain many random terms, it is natural to wonder whether we can invoke the Central Limit Theorem to argue that stochastic lensing can be described in terms of (multivariate) Gaussian distributions.  The answer, unfortunately, is no, because in some of the sums the individual terms have divergent variances.  Previous studies typically sidestepped this issue by using the characteristic function method to compute probability distributions.  That method requires direct and inverse Fourier transforms whose calculations are challenging but feasible in the context of a uniform spatial distribution.  It is not yet clear how easily the characteristic function method can be generalised to arbitrary spatial distributions, so in this paper we adopt an approach that is simpler but still very instructive.  The trouble for the Central Limit Theorem can be ascribed to strong perturbations produced by clumps close (in projection) to an image.  We treat these clumps explicitly by deriving full probability distributions for the \emph{most extreme} perturbations produced by individual clumps.  Once we have isolated the troublemakers in this way, we can apply the Central Limit Theorem to some subset of the remaining clumps; we specifically consider clumps that are projected far from an image.  We develop formal methods to treat these ``local'' and ``long-range'' regimes, which allow us (1) to draw general conclusions about how substructure lensing depends on the mass function and spatial distribution of clumps, and (2) to obtain analytic results that will serve as useful limits for an eventual complete theory of substructure lensing.

Let us be clear from the outset that we make certain simplifying assumptions in the analysis presented here.  As in all previous formal work on stochastic lensing, we approximate the clumps as point masses, and we assume the clumps are independent and identically distributed.  We gear the discussion toward applications that involve compact (essentially point-like) sources.  We compute probability distributions for lensing quantities at fixed points in the image plane (rather than for a fixed source position), because this is conceptually straightforward and also relevant for interpreting observed lenses in which the image positions are known.  We address the utility and validity of these assumptions as they arise in the analysis, and mention possible extensions to the current work at the end of the paper.

%=====================================================================
\section{Fundamentals}
\label{sec:fundamentals}
%=====================================================================

%=====================================================================
\subsection{Basic theory}
\label{sec:basic}

All key lensing quantities can be derived from the lensing potential, $\phi(\vx)$, which is a scaled version of the two-dimensional Newtonian gravitational potential and is given by the Poisson equation $\nabla^2\phi = 2\kappa$.  (See \citealt{SEF}, \citealt{PLW}, and \citealt{saasfee} for general reviews of strong lens theory.)  Here $\kappa = \Sigma/\Sigcr$ is the projected surface mass density in units of the critical density for lensing, $\Sigcr = (c^2 D_s)/(4\pi G D_l D_{ls})$, where $D_l$, $D_s$, and $D_{ls}$ are angular diameter distances to the lens, to the source, and from the lens to the source, respectively.  The lensing potential can be used to determine the time delay, defined to be the excess travel time compared with a light ray that travels directly from the source to the lens:
\begin{equation} \label{eq:tdel}
  \tau(\vx;\vu) = \frac{1+z_l}{c}\ \frac{D_l D_s}{D_{ls}}
    \left[ \frac{1}{2}\left|\vx-\vu\right|^2 - \phi(\vx) \right]
\end{equation}
where $\vx$ and $\vu$ are the angular positions of the image and source, respectively, and $z_l$ is the redshift of the lens.  By Fermat's principle, images form at stationary points of the time delay surface; the condition $\nabla_x \tau(\vx;\vu) = 0$ immediately yields the familiar lens equation,
\begin{equation} \label{eq:lens}
  \vu = \vx - \va(\vx)
\end{equation}
where $\va = \nabla\phi$ is the deflection vector.  The distortion of a (small) image is governed by the magnification tensor,
\begin{eqnarray} \label{eq:Mmat}
  \Mmat = \left(\frac{\partial\vu}{\partial\vx}\right)^{-1}
  = \left[\begin{array}{cc}
    1 - \phi_{xx} & - \phi_{xy} \\
    - \phi_{xy} & 1 - \phi_{yy}
    \end{array}\right]^{-1}
  = \left[\begin{array}{cc}
    1 - \kappa - \gamma_c & - \gamma_s \\
    - \gamma_s & 1 - \kappa + \gamma_c
    \end{array}\right]^{-1}
\end{eqnarray}
where subscripts on $\phi$ denote partial derivatives, e.g., $\phi_{xy} = \partial^2\phi/\partial x\,\partial y$.  In the second expression we have identified $\kappa = (\phi_{xx}+\phi_{yy})/2$ from the Poisson equation; this quantity is referred to as convergence because it describes the focusing of light rays.  We have also defined
\begin{equation}
  \gamma_c = \frac{1}{2}\left(\phi_{xx}-\phi_{yy}\right)
  \qquad\mbox{and}\qquad
  \gamma_s = \phi_{xy}
\end{equation}
These quantities are referred to as shear because they describe how an image is stretched by lensing.

%=====================================================================
\subsection{Smooth and lumpy components}
\label{sec:smooth+lumpy}

The potential ($\phi$), deflection components ($\alpha_x$ and $\alpha_y$), convergence ($\kappa$), and shear components ($\gamma_c$ and $\gamma_s$) are all linear quantities, meaning that each one can be written as a sum of contributions from different components of the mass model.  In substructure lensing we write the full lens potential as
\begin{equation}
  \phi = \phi\smth + \phi\sub
\end{equation}
The term $\phi\smth$ represents the contribution from a smooth component that contains the majority of the mass of the galaxy.\footnote{The superscript here denotes the smooth component and should not be confused with an exponent.}  The term $\phi\sub$ represents the potential from the substructure (which is itself a sum of contributions from many individual clumps; see \refsec{complex}).  Since the deflection components, convergence, and shear components are linear in $\phi$, they can also be written as sums of smooth and lumpy terms: for example, we can write the full deflection vector as $\va = \va\smth + \va\sub$, and likewise for the convergence and shear.

Any lensing observable that is based on image positions or time delays formally involves the time delay and lens equations (eqs.~\ref{eq:tdel} and \ref{eq:lens}), which in turn depend on the substructure potential and deflection.  Thus, for position and time delay observables it is sufficient to focus the theory on understanding probability distributions for $\phi\sub$ and $\va\sub$.

Lensing observables that involve the shapes and brightnesses of images, by contrast, involve the magnification tensor and the scalar magnification, which are nonlinear.  To discuss the magnification, it is convenient to define second derivative matrices:
\begin{equation} \label{eq:Gmat}
  \Gmat\smth = \left[\begin{array}{cc}
    \phi\smth_{xx} & \phi\smth_{xy} \\
    \phi\smth_{xy} & \phi\smth_{yy}
    \end{array}\right]
  = \left[\begin{array}{cc}
    \kappa\smth + \gamma\smth_c & \gamma\smth_s \\
    \gamma\smth_s & \kappa\smth - \gamma\smth_c
    \end{array}\right]
  \quad\mbox{and}\quad
  \Gmat\sub = \left[\begin{array}{cc}
    \phi\sub_{xx} & \phi\sub_{xy} \\
    \phi\sub_{xy} & \phi\sub_{yy}
    \end{array}\right]
  = \left[\begin{array}{cc}
    \kappa\sub + \gamma\sub_c & \gamma\sub_s \\
    \gamma\sub_s & \kappa\sub - \gamma\sub_c
    \end{array}\right]
\end{equation}
The magnification tensor associated with the smooth component alone is (cf.\ eq.~\ref{eq:Mmat})
\begin{equation} \label{eq:Msmth}
  \Mmat\smth = (\Imat-\Gmat\smth)^{-1}
\end{equation}
where $\Imat$ is the $2\times2$ identity matrix.  The scalar magnification associated with the smooth component is
\begin{equation} \label{eq:musmth}
  \mu\smth = \det\Mmat\smth = \frac{1}{(1-\kappa\smth)^2-(\gamma\smth_c)^2-(\gamma\smth_s)^2}
\end{equation}
The magnification tensor for the full model is
\begin{equation} \label{eq:Msub}
  \Mmat = (\Imat-\Gmat\smth-\Gmat\sub)^{-1}
  = \left\{(\Imat-\Gmat\smth)[\Imat-(\Imat-\Gmat\smth)^{-1}\Gmat\sub]\right\}^{-1}
  = \left(\Imat-\Mmat\smth\Gmat\sub\right)^{-1} \Mmat\smth
\end{equation}
Here we have used matrix identities to simplify the expression, and recognised the factor $\Mmat\smth = (\Imat-\Gmat\smth)^{-1}$.  The simplified expression is useful when we write the scalar magnification associated with the full model:
\begin{equation} \label{eq:musub}
  \mu = \det\Mmat = \frac{\det\Mmat\smth}{\det(1-\Mmat\smth\Gmat\sub)} = \frac{\mu\smth}{\det(1-\Mmat\smth\Gmat\sub)}
\end{equation}
It is interesting now to consider the case when the convergence and shear from the substructure are small: $\kappa\sub$, $\gamma\sub_c$, and $\gamma\sub_s$ are all $\order{\epsilon}{}$ where $\epsilon \ll 1$.  While this may not be true in general,\footnote{Indeed, flux ratio anomalies indicate that substructure can have order unity effects on magnifications; B2045+265 provides a good example \citep{fassnacht-2045,cuspreln}.} the perturbative analysis can still offer useful insight.  If we define the magnification perturbation relative to the smooth model, $\dmu = \mu - \mu\smth$, we can write the fractional magnification perturbation as
\begin{equation} \label{eq:dmusub}
  \frac{\dmu}{\mu\smth} = \frac{1}{\det(1-\Mmat\smth\Gmat\sub)} - 1
  = \tr(\Mmat\smth\Gmat\sub) + \left[ \tr(\Mmat\smth\Gmat\sub)^2 - \det(\Mmat\smth\Gmat\sub) \right] + \order{\epsilon}{3}
\end{equation}
The final expression represents a Taylor series expansion in $\epsilon$, using the fact that each component of $\Gmat\sub$ is $\order{\epsilon}{}$ so $\tr(\Mmat\smth\Gmat\sub) \sim \order{\epsilon}{}$, while $\det(\Mmat\smth\Gmat\sub) \sim \order{\epsilon}{2}$ since the matrices are $2\times2$.  (The series expansion would have a different form if the matrices had a different dimensionality.)  In the case of point mass clumps, $\kappa\sub=0$ so we can write the fractional magnification perturbation as
\begin{equation}
  \frac{\dmu}{\mu\smth} = 2\mu\smth \left(\gamma\smth_c\gamma\sub_c+\gamma\smth_s\gamma\sub_s\right)
  + \mu\smth \left[ (\gamma\sub_c)^2 + (\gamma\sub_s)^2 + 4\mu\smth\left(\gamma\smth_c\gamma\sub_c+\gamma\smth_s\gamma\sub_s\right)^2 \right]
  + \order{\epsilon}{3}
\end{equation}
We can take one more step by defining pseudo-polar coordinates for the shear:
\begin{equation}
  \gamma\smth_c = \gamma\smth \cos 2\theta\smth_\gamma
  \quad\mbox{and}\quad
  \gamma\smth_s = \gamma\smth \sin 2\theta\smth_\gamma
\end{equation}
Here $\gamma\smth$ and $\theta\smth_\gamma$ represent the amplitude and direction of the shear from the smooth component; the factor of 2 enters the trig functions because shear is symmetric through a $180^\circ$ rotation (and this is why we call these ``pseudo-polar'' coordinates).  We can likewise define $\gamma\sub$ and $\theta\sub_\gamma$ to be the amplitude and direction of the shear from substructure.  We can then use the identity
\begin{equation}
  \gamma\smth_c\gamma\sub_c + \gamma\smth_s\gamma\sub_s = \gamma\smth \gamma\sub \cos 2(\theta\smth_\gamma-\theta\sub_\gamma)
\end{equation}
to write the fractional magnification perturbation as
\begin{equation}
  \frac{\dmu}{\mu\smth} = 2\mu\smth \gamma\smth \gamma\sub \cos 2(\theta\smth_\gamma-\theta\sub_\gamma)
  + \mu\smth (\gamma\sub)^2 \left[ 1 + 4 \mu\smth (\gamma\smth)^2 \cos^2 2(\theta\smth_\gamma-\theta\sub_\gamma) \right]
  + \order{\epsilon}{3}
\end{equation}
There are several interesting conclusions to draw from this analysis.  First, even the \emph{fractional} magnification perturbation is proportional to $\mu\smth$; and one of the second-order terms is actually quadratic in $\mu\smth$.  A given substructure shear will therefore cause a stronger flux ratio anomaly in a more magnified image.  The factor of $\mu\smth$ means the sign of the fractional magnification perturbation depends on the parity of the image (at lowest order).  The sign is also affected, though, by the $\cos 2(\theta\smth_\gamma-\theta\sub_\gamma)$ factor.  To the extent that the direction of the substructure shear is correlated with the direction of the smooth shear, the $\cos$ factor will be more likely to have a positive sign, and so positive-parity images ($\mu_0>0$) will tend to be brightened by substructure ($\dmu/\mu_0>0$) whereas negative-parity images ($\mu_0<0$) will tend to be dimmed ($\dmu/\mu_0<0$).  If the directions of the smooth and substructure shears are uncorrelated, however, the fractional magnification perturbation will be equally likely to have either sign and so the parity dependence will be masked.  Another consequence of the cosine factor is that the fractional magnification perturbation is only sensitive to the component of the substructure shear that is ``aligned'' with the smooth shear (for shear, ``aligned'' means either parallel or perpendicular).  The ``cross'' component of the substructure shear (i.e., the component oriented at $45^\circ$ with respect to the smooth shear) does not affect the magnification perturbation.  All of these conclusions apply only at lowest order in the substructure shear, so they are not fully general, but they are interesting nonetheless.

Even for arbitrarily large substructure effects, \refeq{dmusub} shows that the magnification perturbation can be written in terms of the convergence and shear from substructure.  More generally, we see from the discussion in this subsection that all of the key lensing observables can be written in terms of the substructure quantities $\Phi\sub \equiv (\phi\sub,\alpha\sub_x,\alpha\sub_y,\kappa\sub,\gamma\sub_c,\gamma\sub_s)$; thus, if we can determine the probability density for $\Phi\sub$ (actually, the joint probability density for $\Phi\sub$ at all image positions) we will have all the information we need to describe lensing observables.  Because of the linearity, $p(\Phi\sub)$ does not depend on the smooth component, which means that we do not need to discuss the smooth component further.  In the remainder of this paper, $(\phi,\alpha_x,\alpha_y,\kappa,\gamma_c,\gamma_s)$ always refer to substructure quantities, and we drop the superscript ``s'' to simplify the notation.

%=====================================================================
\subsection{Complex notation}
\label{sec:complex}

Lens theory naturally operates in two dimensions, and it is customary to treat positions and deflections as real vectors in a two-dimensional plane.  However, it is also possible to think of that plane as the complex plane by combining coordinates $x$ and $y$ into the complex number
\begin{equation}
  w = x + i y
\end{equation}
We can also define the complex deflection and shear,
\begin{equation}
  \ahat = \alpha_x + i \alpha_y
  \quad\mbox{and}\quad
  \ghat = \gamma_c + i \gamma_s
\end{equation}
Note that we use tildes to identify these as complex quantities which are distinct from their scalar amplitudes, $\alpha = |\ahat|$ and $\gamma = |\ghat|$.  In various circumstances the complex notation can simplify lens theory \citep[e.g.,][]{bourassa1,bourassa2,witt,rhie,PLW,an2005,an2007,an2006,khavinson1,khavinson2}, and we shall find it to be useful.  In complex variables, the potential, deflection, shear, and convergence at position $w$ created by a point mass clump $m_i$ at position $w_i$ have the form:
\begin{eqnarray}
  \phi_i(w)   &=&  \frac{m_i}{\pi} \ln\frac{|w-w_i|}{a}     \label{eq:poti} \\
  \ahat_i(w)  &=&  \frac{m_i}{\pi}\ \frac{1}{w^*-w_i^*}     \label{eq:defi} \\
  \ghat_i(w)  &=& -\frac{m_i}{\pi}\ \frac{1}{(w^*-w_i^*)^2} \label{eq:shri} \\
  \kappa_i(w) &=&  m_i\,\delta_D(w-w_i)                     \label{eq:kapi}
\end{eqnarray}
where the asterisk denotes complex conjugation, and $\delta_D$ is the Dirac delta function.  Also, $m_i = M_i/\Sigcr$ is the physical mass of the clump scaled by the critical surface density for lensing, so it is a quantity with dimensions of area.  In the potential, $a$ is an arbitrary length scale that can be used to set the zeropoint of the potential.  We discuss the zeropoint as needed below.  The total substructure potential at position $w$ due to all clumps can then be written as
\begin{equation} \label{eq:phisub}
  \phi(w) = \sum_i \phi_i(w)
\end{equation}
and there are similar expressions for the total deflection, shear, and convergence.

At times it is useful to convert from the complex notation into conventional polar coordinates.  If we write $w = r\,e^{i\theta}$ then $r$ and $\theta$ have their usual relation to $x$ and $y$.  We also consider polar coordinates centred on the position of an image by writing $w' = w - w_\img = r'\,e^{i\theta'}$.  If the image itself has polar coordinates $(\rimg,\timg)$, the two polar coordinate systems are related by
\begin{equation} \label{eq:polar}
  r^2 = \rimg^2 + r'^2 + 2 \rimg r' \cos(\theta'-\timg)
  \qquad\mbox{and}\qquad
  r'^2 = \rimg^2 + r^2 - 2 \rimg r \cos(\theta-\timg)
\end{equation}

%=====================================================================
\subsection{Clump population}
\label{sec:clumppop}

In this paper we use point mass clumps, so each clump has a position, $w_i$, and scaled mass, $m_i$.  The point mass approximation is valid for any spherical clump that does not overlap the line of sight, because the gravitational field outside such a clump equivalent to that of a point mass clump.  There will be some correction for clumps that overlap the line of sight, but we leave for future work the development of a general theory that includes spatially extended clumps.  (See \citealt{rozo} for an analysis of magnification perturbations that can handle extended clumps.)

The statistical properties of the clump population are specified by the joint probability density $p(w_1,m_1,w_2,m_2,\ldots)$.  We assume the clumps are independent and identically distributed, so the joint probability density can be written as a product of probabilities densities for individual clumps:
\begin{equation}
  p(w_1,m_1,w_2,m_2,\ldots) = p(w_1,m_1) \times p(w_2,m_2) \times \ldots
\end{equation}
This approach neglects possible correlations among clumps, but even if such correlations exist in three dimensions they will be suppressed to some extent by projection effects in lensing.  Correlations among clumps would further complicate the theory and so we defer them to future work.  We also assume the positions and masses are separable, so we can write
\begin{equation}
  p(w_i,m_i) = p_w(w_i)\, p_m(m_i)
\end{equation}
Here $p_w(w)$ is the spatial probability density, defined such that $p_w(w)\,dw$ is the probability that a given clump is found within some area element $dw$ around position $w$ (in the complex plane).  Also, $p_m(m)$ is the mass probability density such that $p_m(m)\,dm$ is the probability that a given clump has mass between $m$ and $m+dm$; the equivalent clump mass function is $dN/dm = N\,p_m(m)$.  Numerical simulations suggest that mass and position may not be fully separable; clumps in different mass ranges tend to have somewhat different spatial distributions due to effects such as tidal stripping and disruption \citep[e.g.,][]{ghigna,delucia,nagai}.  Such an effect could be accounted for in our formalism by dividing the clump population into several sub-populations by mass and using a different spatial distribution for each.

To gain a more intuitive understanding of the spatial probability density, consider the mean surface mass density in substructure:
\begin{equation} \label{eq:p2kap}
  \ksub(w) \equiv \avg{\kappa(w)} = N \mbar p_w(w)
\end{equation}
Here ``mean'' refers to averaging over many realizations of the substructure, in a sense that is formalised in \refapp{stats} (see eq.~\ref{eq:genavg}, with $f_i \to \kappa_i(w) = m_i\,\delta_D(w-w_i)$ from eq.~\ref{eq:kapi}).  Later in the paper we use this relation to replace the abstract quantity $p_w(w)$ with the more physical quantity $\ksub(w)$ in spatial integrals.

%=====================================================================
\subsection{Sample models}
\label{sec:sample}

Even as we attempt to build a general theory of lensing with stochastic substructure, there are times when it is helpful to consider specific sample clump populations.  For the spatial distribution, one simple model we consider is a uniform distribution,
\begin{equation} \label{eq:eg-uni}
  p_x(w) = \frac{\ksub}{N \mbar}
\end{equation}
defined over some large but finite area such that $\int p_w(w)\ dw = 1$.  The total number of clumps $N$ is basically interchangeable with the area over which $p_w$ is defined.  As discussed in the Introduction, the spatially uniform case allows us to connect with previous work on stochastic lensing.  A second spatial model we consider is a power law distribution,
\begin{equation} \label{eq:eg-pow1}
  p_w(w) \propto |w|^{\eta-2}
  \qquad\mbox{with}\qquad
  0 < \eta \le 2
\end{equation}
The power law index is chosen such that the total mass in substructure within projected radius $r$ scales as $M(r) \propto r^\eta$.  In terms of image-centred polar coordinates we can write
\begin{equation} \label{eq:eg-pow2}
  p_w \propto \left[1 + \frac{r'^2}{\rimg^2} + 2 \frac{r'}{\rimg} \cos\theta'\right]^{(\eta-2)/2}
\end{equation}
While the uniform and power law models do not necessarily match simulated CDM substructure populations in detail, they are useful pedagogical examples and sufficient for our purposes in this paper.

CDM simulations predict that the mass function is a power law over many orders of magnitude, $dN/dm \propto m^\beta$ with $\beta \approx -1.9$ \citep[e.g.,][]{VL-sub,aquarius}.  To understand how substructure lensing depends on clump mass, it is instructive to consider a finite mass range $m_1 \le m \le m_2$.  The mean clump mass is then
\begin{equation}
  \avg{m} = m_1\ \frac{q^{2+\beta}-1}{2+\beta}\ \frac{1+\beta}{q^{1+\beta}-1}
\end{equation}
where $q = m_2/m_1$ is the dynamic range of the mass function.  More generally, other moments of the mass function have the form
\begin{equation}
  \avg{m^n} = m_1^n\ \frac{q^{n+1+\beta}-1}{n+1+\beta}\ \frac{1+\beta}{q^{1+\beta}-1}
\end{equation}
As we shall see, there is one combination of mass moments that arises repeatedly:
\begin{equation} \label{eq:meff}
  \meff \equiv \frac{\avg{m^2}}{\avg{m}}
  = \avg{m}
    \frac{q^{1+\beta}-1}{1+\beta}\ 
    \frac{q^{3+\beta}-1}{3+\beta}
    \left(\frac{2+\beta}{q^{2+\beta}-1}\right)^2
\end{equation}
This quantity has dimensions of mass, and has been referred to in microlensing as the ``effective mass'' \citep[e.g.,][]{refsdal-meff,neindorf}.

We define a fiducial model whose spatial distribution is a power law with $\eta=1$ (corresponding to an isothermal profile), and whose mass function has mean mass $\avg{M} = 10^8\,h^{-1}\,M_\odot$ and dynamic range $q=100$.  We then construct variants with a steeper or shallower radial profile ($\eta=0.5$ or 1.5, respectively); all models are normalised to have $\ksub=0.01$ in the vicinity of the lensed images.  We also construct other mass functions that have different dynamic ranges but are all normalised to have the same effective mass (see \reftab{massfunc}).  Note that the quantitative details of our sample models are not important; what matters is that the examples illustrate the key concepts and scalings derived from our mathematical analysis.

\begin{table}
\caption{Sample clump mass functions.  All have power law index $\beta=-1.9$ and are normalised to have the same effective mass, $\meff=\avg{m^2}/\avg{m}$.  Column 1 gives the dynamic range, $q=m_2/m_1$.  Column 2 gives the lower end of the mass range.  Column 3 gives the mean mass.  The remaining columns give higher moments (taking appropriate powers so each quantity has dimensions of mass).  All masses are given in units of $10^8\,h^{-1}\,M_\odot$.}
\begin{tabular}{rlllll}
\hline
 $q$ & $M_1$  & $\avg{M}$ & $\avg{M^2}^{1/2}$ & $\avg{M^3}^{1/3}$ & $\avg{M^4}^{1/4}$ \\
\hline
   1 & 4.58   & 4.58  & 4.58 & 4.58 & 4.58 \\
   3 & 2.49   & 4.14  & 4.35 & 4.57 & 4.77 \\
  10 & 1.12   & 3.00  & 3.70 & 4.43 & 5.08 \\
  30 & 0.496  & 1.90  & 2.95 & 4.10 & 5.14 \\
 100 & 0.187  & 1.00  & 2.14 & 3.56 & 4.89 \\
 300 & 0.0731 & 0.509 & 1.53 & 2.99 & 4.47 \\
1000 & 0.0251 & 0.225 & 1.02 & 2.39 & 3.90 \\
\hline
\label{tab:massfunc}
\end{tabular}
\end{table}

We consider a cosmology with $\Omega_M=0.274$ and $\Omega_\Lambda=0.726$ \citep{komatsu}.  We set the lens and source redshifts to $z_l=0.31$ and $z_s=1.722$ \citep[as for PG~1115+080;][]{weymann-1115,henry-1115}.  Then the cosmological distances are $D_l=662\,h^{-1}\,\mbox{Mpc}$, $D_s=1248\,h^{-1}\,\mbox{Mpc}$, and $D_{ls}=930\,h^{-1}\,\mbox{Mpc}$, and the critical density for lensing is $\Sigcr = 3378\,h\,M_\odot\,\mbox{pc}^{-2}$ or equivalently $\Sigcr = 3.48 \times 10^{10}\,h^{-1}\,M_\odot\,\mbox{arcsec}^{-2}$.

%=====================================================================
\section{Local analysis}
\label{sec:local}
%=====================================================================

In the Introduction we noted that the probability distributions for the substructure shear and deflection have divergent variances.  In this section we analyze the ``heavy tails'' that cause the variance to diverge, by computing the probability distributions for the most extreme shear, deflection, and variance.  (We do not consider the convergence because it is zero for point mass clumps; see eq.~\ref{eq:kapi}.)  We seek to understand how the mass function and spatial distribution of the clump population affect the strongest substructure perturbations.  For the analysis in this section we focus on one particular image and compute one-point statistics.\footnote{In this section we do not consider two-point statistics because we expect the local effects for different images to be (largely) independent: the clump that produces the largest shear for one image is not likely to be the clump that produces the largest shear for another image.}  We work in coordinates centred on that image, $w' = w_i-w_\img$, which allows us to write the amplitudes of the shear and deflection from clump $i$ as
\begin{equation}
  \gamma_i = |\ghat_i| = \frac{m_i}{\pi |w'_i|^2}
  \qquad\mbox{and}\qquad
  \alpha_i = |\ahat_i| = \frac{m_i}{\pi |w'_i|}
\end{equation}
and the potential from clump $i$ as
\begin{equation}
  \phi_i = \frac{m_i}{\pi} \ln\frac{|w'_i|}{a}
\end{equation}
where again $a$ is a length scale that sets the zeropoint of the potential.  

%=====================================================================
\subsection{Uniform spatial distribution}

We begin with the simple case of a uniform spatial distribution to introduce the methods.  From \refeq{eg-uni}, the spatial probability distribution is $p_w(w) = \ksub/N \mbar$.  When $N$ is finite the area over which $p_w(w)$ is defined is also finite (albeit large), but we will not be too concerned with the boundary because we will eventually take the limit $N \to \infty$.

First consider the shear.  We seek to determine the probability distribution for the largest shear produced by any individual clump.  Before we can do that, we need to find the probability that the shear from a given clump $i$ is bigger than $\gamma$:
\begin{equation} \label{eq:Ploc1g}
  P_i(>\!\gamma) = \int_{\frac{m_i}{\pi |w'_i|^2}>\gamma}  p_w(w'_i)\ p_m(m_i)\ dw'_i\ dm_i
  = \frac{1}{N\mbar} \int dm\ p_m(m) \int d\theta' \int_{0}^{\left(\frac{m}{\pi\gamma}\right)^{1/2}} dr'\ r'\ \ksub
  = \frac{\ksub}{N\gamma}
\end{equation}
Here we have written the expression for the total probability in the region where $\gamma_i > \gamma$, then plugged in for $p_w(w')$ and written the $w'$ integral in terms of polar coordinates (using $w' = r'\,e^{i\theta'}$), and finally evaluated the integrals.  Strictly speaking, this analysis is not valid to arbitrarily small values of $\gamma$, because small shears can only be produced by clumps that are far from the image, and the clump domain has some finite extent.  Put another way, $P_i(>\!\gamma)$ cannot exceed unity, so clearly this expression is valid only for $\gamma \ge \ksub/N$.  This detail will become immaterial when we take the limit $N \to \infty$.

Now, the probability that the shear from clump $i$ is smaller than $\gamma$ is obviously $1-P_i(>\!\gamma)$.  Since the clumps are independent, the probability that the shears from all clumps are smaller than $\gamma$ is obtained by taking the product of all the individual clump probabilities:
\begin{equation} \label{eq:Ploc2g}
  P_{\rm all}(<\!\gamma) = \prod_{i=1}^{N} \bigl[1-P_i(>\!\gamma)\bigr]
  = \left(1-\frac{\ksub}{N\gamma}\right)^{N}
  \to \exp\left(-\frac{\ksub}{\gamma}\right) \quad \mbox{ for } N \to \infty
\end{equation}
This is the cumulative probability distribution for the largest shear amplitude.  Notice that the clump mass function dropped out of the analysis: the probability distribution for the largest shear does not depend on how the clumps are distributed in mass, when the spatial distribution is uniform \citep[also see][]{schneider-shear,alberto2}.

Next we consider the probability distribution for the largest deflection amplitude.  The argument proceeds just as before.  The probability that the deflection from a given clump $i$ is bigger than $\alpha$ is:
\begin{equation} \label{eq:Ploc1a}
  P_i(>\!\alpha) = \int_{\frac{m_i}{\pi |w'_i|}>\alpha}  p_w(w'_i)\ p_m(m_i)\ dw'_i\ dm_i
  = \frac{1}{N\mbar} \int dm\ p_m(m) \int d\theta' \int_{0}^{\frac{m}{\pi\alpha}} dr'\ r'\ \ksub
  = \frac{\ksub}{N\pi\alpha^2}\,\frac{\avg{m^2}}{\mbar}
\end{equation}
The probability that all deflections are smaller than $\alpha$ is then:
\begin{equation} \label{eq:Ploc2a}
  P_{\rm all}(<\!\alpha) = \left(1-\frac{\ksub}{N\pi\alpha^2}\,\frac{\avg{m^2}}{\mbar}\right)^{N}
  \to \exp\left(-\frac{\ksub}{\pi\alpha^2}\,\frac{\avg{m^2}}{\mbar}\right)
    \quad \mbox{ for } N \to \infty
\end{equation}
This is the cumulative probability distribution for the largest deflection amplitude.  Here we see that the probability distribution for the largest deflection does depend on the clump mass function, but only through the combination of moments known as effective mass, $\meff = \avg{m^2}/\avg{m}$.  This is consistent with previous results showing that the deflection probability distribution depends on the mass function through $\meff$ when the spatial distribution is uniform \citep{katz,neindorf}.

Finally we turn to the potential.  The only difference in this case is that the potential \emph{increases} with the distance of the clump from the image, so the inequalities are reversed.  Specifically, we want to determine the probability distribution for the ``strongest'' (i.e., most negative) potential.  The probability that the potential from a given clump $i$ is less than $\phi$ is:
\begin{eqnarray} \label{eq:Ploc1p}
  P_i(<\!\phi) &=& \int_{\frac{m_i}{\pi}\ln\frac{|w'_i|}{a}<\phi} p_w(w'_i)\ p_m(m_i)\ dw'_i\ dm_i
  = \frac{1}{N\mbar} \int dm\ p_m(m) \int d\theta' \int_{0}^{a e^{\pi\phi/m}} dr'\ r'\ \ksub \\
  &=& \frac{\ksub \pi a^2}{N \mbar} \int e^{2\pi\phi/m}\ p_m(m)\ dm \nonumber
\end{eqnarray}
The probability that all potentials are larger than $\phi$ is then (for $N\to\infty$):
\begin{equation} \label{eq:Ploc2p}
  P_{\rm all}(>\!\phi) = \exp\left(-\frac{\ksub \pi a^2}{\mbar} \int e^{2\pi\phi/m}\ p_m(m)\ dm\right)
\end{equation}
This is the cumulative probability distribution for the strongest potential.  In this case we cannot express the mass integral in terms of any simple moment of the mass function.

%=====================================================================
\subsection{General case}

For an arbitrary spatial distribution, the analogues of eqs.~(\ref{eq:Ploc1g}), (\ref{eq:Ploc1a}), and (\ref{eq:Ploc1p}) are
\begin{eqnarray}
  P_i(>\!\gamma) &=& \int_{\frac{m_i}{\pi |w'_i|^2}>\gamma} p_w(w'_i)\ p_m(m_i)\ dw'_i\ dm_i
  = \frac{1}{N\mbar} \int dm\ p_m(m) \int d\theta'
    \int_{0}^{\left(\frac{m}{\pi\gamma}\right)^{1/2}} dr'\ r'\ \ksub(r',\theta')
  \label{eq:Ploc3g} \\
  P_i(>\!\alpha) &=& \int_{\frac{m_i}{\pi |w'_i|}>\alpha}  p_w(w'_i)\ p_m(m_i)\ dw'_i\ dm_i
  = \frac{1}{N\mbar} \int dm\ p_m(m) \int d\theta'
    \int_{0}^{\frac{m}{\pi\alpha}} dr'\ r'\ \ksub(r',\theta')
  \label{eq:Ploc3a} \\
  P_i(<\!\phi) &=& \int_{\frac{m_i}{\pi}\ln\frac{|w'_i|}{a}<\phi} p_w(w'_i)\ p_m(m_i)\ dw_i\ dm_i
  = \frac{1}{N\mbar} \int dm\ p_m(m) \int d\theta'
    \int_{0}^{a e^{\pi\phi/m}} dr'\ r'\ \ksub(r',\theta')
  \label{eq:Ploc3p}
\end{eqnarray}
Notice that the form of the integral is the same in all three cases; the only difference is the upper limit of the radial integral.  Therefore let us define
\begin{equation} \label{eq:Qdef}
  Q(z) = \int d\theta' \int_{0}^{z} dr'\ r'\ \khat(r',\theta')
\end{equation}
where $\khat(r',\theta') = \ksub(r',\theta')/\kimg$ is the mean substructure density at the position specified by $(r',\theta')$ normalised by the value at the image.  Normalising by $\kimg$ means that the function $Q(z)$ is independent of the amount of substructure and depends only on the spatial distribution of the substructure population. Note that for the uniform case, $Q(z) = \pi z^2$.  Now we rewrite eqs.~(\ref{eq:Ploc3g})--(\ref{eq:Ploc3p}) using $Q(z)$, and then repeat the argument that takes us from \refeq{Ploc1g} to \refeq{Ploc2g}, to obtain general expressions for the probability distributions for the local shear, deflection, and potential (in the limit $N\to\infty$):
\begin{eqnarray}
  P(<\!\gamma) &=& \exp\left[ -\frac{\kimg}{\mbar}
    \int Q\left(\sqrt{\frac{m}{\pi\gamma}}\right)\ p_m(m)\ dm\right]
    \label{eq:Ploc4g}\\
  P(<\!\alpha) &=& \exp\left[ -\frac{\kimg}{\mbar}
    \int Q\left(\frac{m}{\pi\alpha}\right)\ p_m(m)\ dm\right]
    \label{eq:Ploc4a}\\
  P(>\!\phi) &=& \exp\left[ -\frac{\kimg}{\mbar}
    \int Q\left(a e^{\pi\phi/m}\right)\ p_m(m)\ dm\right]
    \label{eq:Ploc4p}
\end{eqnarray}
One conclusion we can draw from these expressions is the general form of the tails of the probability distributions for shear and deflection.  At lowest order in $z$ we have $Q(z) \propto z^2$ which immediately yields
\begin{eqnarray}
  p(\gamma) = \frac{dP(<\!\gamma)}{d\gamma} \propto \gamma^{-2} \qquad(\gamma\to\infty) \\
  p(\alpha) = \frac{dP(<\!\alpha)}{d\alpha} \propto \alpha^{-3} \qquad(\alpha\to\infty)
\end{eqnarray}
(Note that these are one-dimensional probability distributions for the shear and deflection amplitudes, not two-dimensional distributions for the full shear and deflection.)  These scalings have been derived before for a uniform spatial distribution of equal-mass clumps \citep{nityananda,katz,schneider-shear,alberto1,alberto2}, but now we see that they are quite general.  This is not surprising: since large shears and deflections can only be produced by clumps in the vicinity of an image, the tails of the shear and deflection distributions cannot be very sensitive to the global population of clumps.

%=====================================================================
\subsection{Power law spatial distribution}

So far we have examined a uniform spatial distribution, which is simplistic, and the arbitrary case, which yielded expressions that are fully general but not especially enlightening.  We now split the difference by considering a power law spatial distribution, which is still somewhat simplified but certainly better than the uniform case, and still tractable.  From \refeq{eg-pow2} we can write the spatial factor that appears in $Q(z)$ as
\begin{equation}
  \khat(r',\theta') = \left(1 + \frac{r'^2}{\rimg^2} + 2 \frac{r'}{\rimg} \cos\theta'\right)^{(\eta-2)/2}
\end{equation}
For general $\eta$ we cannot evaluate the integrals in $Q(z)$ analytically.  However, we can make a Taylor series expansion in $z$ that gives a useful approximation in the limit of a large shear or deflection.  The series expansion for $Q(z)$ is
\begin{equation}
  Q(z) = \pi z^2 \left[ 1
  + \frac{(2-\eta)^2}{8} \left(\frac{z}{\rimg}\right)^2
  + \frac{(2-\eta)^2 (4-\eta)^2}{192} \left(\frac{z}{\rimg}\right)^4
  + \frac{(2-\eta)^2 (4-\eta)^2 (6-\eta)^2}{9216} \left(\frac{z}{\rimg}\right)^6
  + \order{\left(\frac{z}{\rimg}\right)}{8} \right]
\end{equation}
We use this expression in \refeqs{Ploc4g}{Ploc4a}, and recognise that the mass integrals yield mass moments of the form $\int m^n\,p_m(m)\,dm = \avg{m^n}$.  This yields series expansions for the probability distributions for the strongest shear and deflection:
\begin{eqnarray}
  P(<\!\gamma) &=& 1 - \frac{\kimg}{\gamma}
    + \frac{\kimg}{\gamma^2}\left[\frac{\kimg}{2}-\frac{\avg{m^2}}{\avg{m}}\frac{(\eta-2)^2}{8\pi\rimg^2}\right]
    + \order{\gamma}{-3} \label{eq:Ploc5g} \\
  P(<\!\alpha) &=& 1 - \frac{\avg{m^2}}{\avg{m}} \frac{\kimg}{\pi\alpha^2}
    + \frac{\kimg}{2\pi^2\alpha^4}\left[\frac{\avg{m^2}^2}{\avg{m}^2} \kimg
      - \frac{\avg{m^4}}{\avg{m}} \frac{(\eta-2)^2}{4\pi\rimg^2}\right]
    + \order{\alpha}{-6} \label{eq:Ploc5a}
\end{eqnarray}
Notice that to lowest order (i.e., for large local shear and deflection), the shear distribution is independent of the mass function while the deflection distribution depends on $\meff = \avg{m^2}/\avg{m}$, and neither is sensitive to the spatial distribution of substructure (i.e., to $\eta$).  This makes sense physically, because large local shears and deflections can only be created by clumps relatively close to the image, and so the tail of the probability distribution depends only on the local abundance of substructure (i.e., $\kimg$).  The spatial distribution enters through higher-order terms, in combination with other moments of the mass function ($\avg{m^2}/\avg{m}$ for the shear, and $\avg{m^4}/\avg{m}$ for the deflection).  Thus, we find that the shear and deflection distributions are formally sensitive to the spatial distribution of substructure along with various moments of the mass function. In practice, however, those sensitivities are important only at relatively low values of the local shear or deflection.

We do not attempt a similar analysis for the potential because it is not possible to express the integral over the mass function in terms of simple mass moments.  The implication is that probability distribution for the local potential is sensitive to the full shape of the mass function.

%=====================================================================
\subsection{Examples}

We now use the sample models discussed in \refsec{sample} to illustrate the results from this section.  We emphasise that the quantitative results presented here depend on our choice of sample models and should not be taken as detailed predictions; our goal with these examples is only to illuminate the concepts drawn from our formal analysis.

\begin{figure}
\centerline{\includegraphics[width=17.0cm]{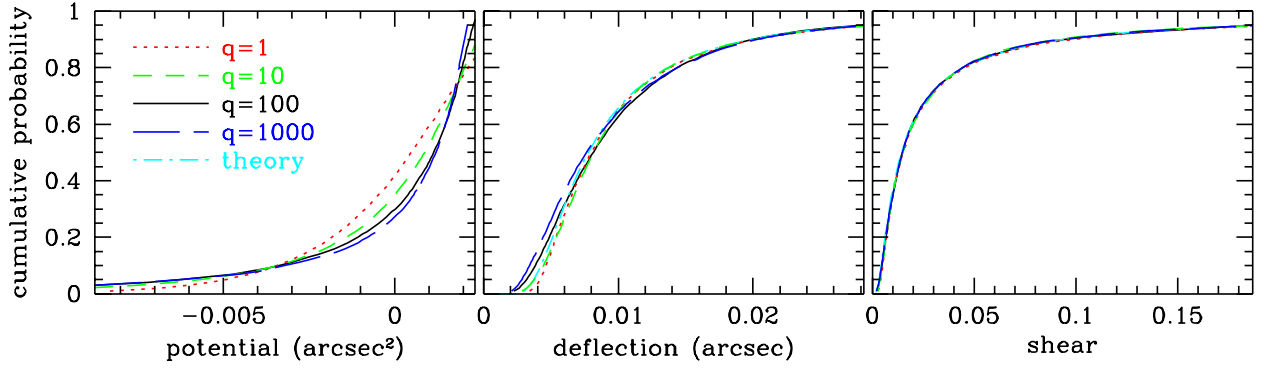}}
\caption{
Cumulative probability distributions for the ``local'' (i.e., most extreme) potential, deflection amplitude, and shear amplitude.  In the text we computed the potential distribution as $P(>\!\phi)$, but here we plot $P(<\!\phi) = 1-P(>\!\phi)$ for consistency with the deflection and shear plots.  The different curves correspond to different mass functions, all normalised to have the same effective mass, $\meff=\avg{m^2}/\avg{m}$ (see \reftab{massfunc}).  The spatial distribution of clumps is a power law with $\eta=1$ and $\ksub=0.01$ at the position of the image we are examining.  In the deflection and shear panels, the cyan dot-dash curves show the theoretical predictions from \refeqs{Ploc2a}{Ploc2g}.  The zeropoint of the potential is chosen so the mean of each potential distribution is zero.
}\label{fig:meff-loc}
\end{figure}

\begin{figure}
\centerline{\includegraphics[width=17.0cm]{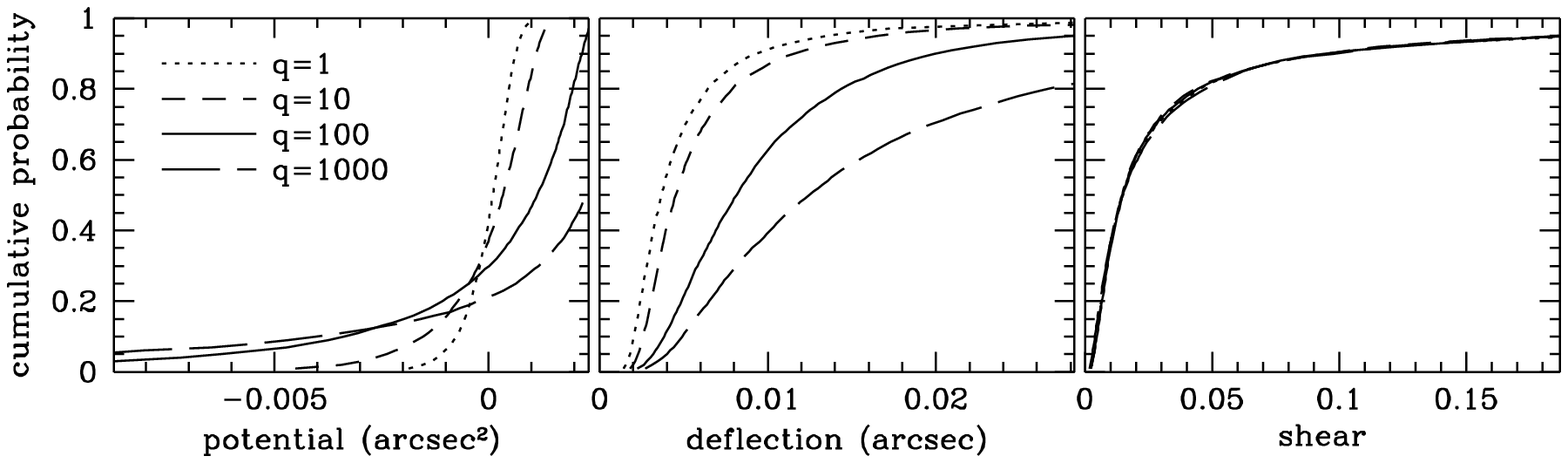}}
\caption{
Similar to \reffig{meff-loc}, but the mass functions are normalised to have the same mean mass, $\avg{m}$ (instead of the same effective mass, $\meff$).
}\label{fig:mbar-loc}
\end{figure}

We generate $10^4$ Monte Carlo realizations of clump populations, tabulate the most extreme substructure terms (the smallest potential, the largest deflection amplitude, and the largest shear amplitude) from each one, and then plot the resulting probability distributions.  For example, \reffig{meff-loc} shows results from simulations with different clump mass functions, when all the mass functions are normalised to have the same effective mass, $\meff=\avg{m^2}/\avg{m}$.  For comparison, \reffig{mbar-loc} shows the corresponding results when all the mass functions are normalised to have the same mean mass, $\avg{m}$.  In both cases the spatial distribution of clumps is a power law with $\eta=1$ and $\ksub=0.01$ at the position of the image we are examining.

The direct simulations show that the local shear distribution is essentially identical for all mass functions, which is consistent with our theoretical expectations; and we see that \refeq{Ploc2g} matches the simulated shear distribution very well.  When we fix $\meff$, the local deflection distribution is basically independent of the clump mass function at large deflections, and only weakly sensitive to the mass function at small deflections.  By contrast, when we fix $\avg{m}$ the deflection distribution is very sensitive to the mass function.  Clearly it is the effective mass, and not the mean mass, that is the important property of the mass function in terms of predicting the local deflection distribution.  Furthermore, we see that \refeq{Ploc2a} gives a very useful approximation for this distribution, except at small deflections (which are probably not of great interest anyway).  There is no obvious, simple scaling for the local potential distribution, although it does appear that the effective mass is more useful than the mean mass for estimating this distribution.

\begin{figure}
\centerline{\includegraphics[width=17.0cm]{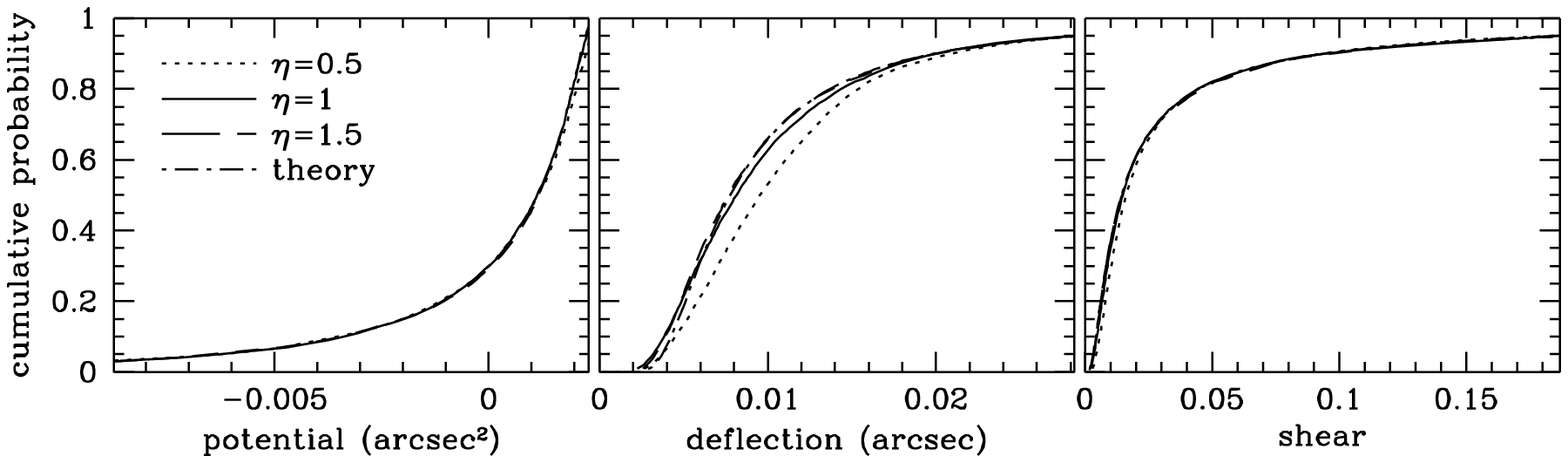}}
\caption{
Similar to \reffig{meff-loc}, but now with different power law indices for the spatial distribution of clumps: $\ksub \propto r^{\eta-2}$.  Results are shown for a mass function with dynamic range $q=100$.
}\label{fig:eta-loc}
\end{figure}

Next we consider varying the spatial distribution of mass clumps, as shown in \reffig{eta-loc}.  The local shear distribution is hardly affected except for small changes at small values of the shear.  The local deflection distribution is somewhat more sensitive to the spatial distribution, especially at modest deflections, although \refeq{Ploc2a} remains a useful approximation.  Both of these results are consistent with our theoretical expectations.  One striking result we did not predict is that the local potential distribution is quite insensitive to the spatial distribution of clumps (at least when the potential zeropoint is chosen such that the mean local potential is zero).  We do not currently have a formal explanation for this empirical result.

\begin{figure}
\centerline{\includegraphics[width=17.0cm]{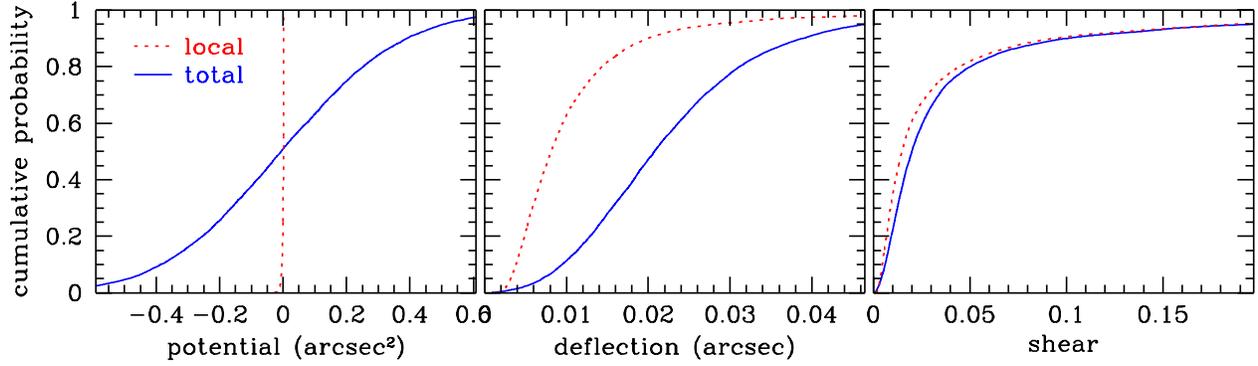}}
\caption{
Comparison of the cumulative probability distributions for the \emph{local} potential, deflection, and shear (red dotted curves) with the corresponding distributions for the \emph{total} potential, deflection, and shear (blue solid curves).  Results are shown for a mass function with dynamic range $q=100$, and a power law spatial distribution with $\eta=1$ and $\ksub=0.01$ at the position of the image.  The zeropoint of the potential is again chosen so the mean of each potential distribution is zero.
}\label{fig:loctot}
\end{figure}

Finally, one thing we can do with the simulations that we could not do with the theory is determine how important the local effects are compared with the total effects due to the full clump population.  \reffig{loctot} shows cumulative probability distributions for the local potential, deflection, and shear alongside the corresponding distributions for the total potential, deflection, and shear.  (For the total deflection and shear, we properly sum the complex terms and then take the amplitude of the final sum.)  We use the fiducial model in which the mass function has a dynamic range $q=100$, and the spatial distribution is a power law with $\eta=1$ and $\ksub=0.01$ at the position of the image.  The local and total shear distributions are not very different, especially at large shears, which indicates that the total shear is dominated by local effects \citep[also see][]{rozo}.  The only difference is when the local shear is small or modest, which leaves room for the combined effects of more distant clumps to become noticeable.  By contrast, there is a more significant difference between the local and total deflection distributions; in other words, there are comparable contributions to the deflection from the clump(s) nearest to the image and from clumps farther away \citep[also see][]{chen-pos}.  As for the potential, the local effects are almost entirely negligible compared with the combined effects of many distant clumps.

These results represent the first stage of our conclusions about how lensing quantities depend on the mass function and spatial distribution of clumps.  They are placed in a broader context in \refsec{discussion} below.

%=====================================================================
\section{Long-range analysis}
\label{sec:long}
%=====================================================================

We now turn attention to clumps that are far from the images.  The potential, deflection, and shear from these clumps can be written as sums of many contributions, all of which are finite, so the Central Limit Theorem suggests that their joint probability density can be approximated as a multivariate Gaussian distribution.  (Again, we do not consider the convergence because it is zero for point mass clumps; see eq.~\ref{eq:kapi}.)  Under this approximation, all we need to know are the mean vector and covariance matrix.  In this section we compute those quantities for the population of clumps outside some radius $R_0$ from the centre of the galaxy.  For simplicity we assume the clump population has circular symmetry, so $\ksub(w)$ is only a function of $|w|$, but we consider arbitrary radial distributions.

Formally, ``mean'' and ``covariance'' refer to averages over many realizations of the clump population.  Details of the averaging process are discussed in \refapp{stats}.  The key expressions are \refeq{avgf} for the mean of any quantity $f$, and \refeq{covfg} for the covariance between two quantities $f$ and $g$.  (Note that we can compute the variance of $f$ as $\cov(f,f)$.)  The expression for the covariance is an approximation that is valid when the number of clumps is large.

Given circular symmetry, the means are trivial: $\avg{\phi}$, $\avg{\ahat}$, and $\avg{\ghat}$ all vanish due to the well-known result that there is no gravitational force inside a circular shell.  Strictly speaking all we can say about the potential inside the shell is that it is constant, but that constant can be absorbed into the zeropoint.  In this section we sidestep the zeropoint by working with the differential potential relative to the origin, i.e., ``$\phi$'' here actually stands for $\phi(w_\img)-\phi(0)$.

To illustrate the covariance calculation let us consider the covariance between the deflections at two images ($\ahat_1$ at position $w_1$, and $\ahat_2$ at position $w_2$).  Using \refeq{covfg} we have:
\begin{eqnarray} \label{eq:covaa}
  \cov(\ahat_1,\ahat_2) &=& \meff \int_{|w_i|>R_0} \frac{1}{\pi(w_1^*-w^*_i)}\ \frac{1}{\pi(w_2-w_i)}\ 
    \ksub(|w_i|)\ dw_i \\
  &=& \frac{\meff}{\pi^2} \int_{0}^{2\pi} d\theta_i \int_{R_0}^{\infty} dr_i\ r_i\ \ksub(r_i)
\left[\frac{e^{i\theta_i}}{r_i}+\frac{w_1^* e^{2i\theta_i}}{r_i^2}+\frac{(w_1^*)^2 e^{3i\theta_i}}{r_i^3}+\ldots\right]
\left[\frac{e^{-i\theta_i}}{r_i}+\frac{w_2 e^{-2i\theta_i}}{r_i^2}+\frac{w_2^2 e^{-3i\theta_i}}{r_i^3}+\ldots\right]
\nonumber\\
  &=& \frac{2\meff}{\pi} K_2 \bigl(1 + K_4 w_1^* w_2 + \ldots\bigr) \nonumber
\end{eqnarray}
In the second line we explicitly consider only clumps outside $R_0$.  We assume that $R_0$ is well outside the images ($r_i > R_0 \gg |w_1|,|w_2|$) and make a Taylor series expansion in $1/r_i$.  We can then evaluate the angular integral and reduce the expression for the covariance to integrals over the spatial distribution of substructure with different radial weighting:
\begin{eqnarray}
  K_2 &=& \int_{R_0}^{\infty} \frac{\ksub(r_i)}{r_i^2}\ r_i\ dr_i \\
  K_4 &=& \frac{1}{K_2} \int_{R_0}^{\infty} \frac{\ksub(r_i)}{r_i^4}\ r_i\ dr_i
\end{eqnarray}
Note that $K_2$ is dimensionless, while $K_4$ has dimensions of 1/length$^2$ and scales as $K_4 \propto R_0^{-2}$.  Assuming the two images are in the vicinity of the lens galaxy's Einstein radius, $|w_1|,|w_2| \sim \Rein$, the second term in parentheses in \refeq{covaa} is of order $(\Rein/R_0)^2$, and any additional terms would be corrections of order $(\Rein/R_0)^4$ and higher.

Now if we consider the potential, deflection, and shear for two different images and assemble them into a (complex) vector $\vv=(\phi_1,\ahat_1,\ghat_1,\phi_2,\ahat_2,\ghat_2)$, then we can use a similar calculation to obtain the full covariance matrix:
\begin{equation} \label{eq:covmat}
  \Cmat = \frac{\meff}{\pi} K_2 \times
\end{equation}
$$
\left[\begin{array}{llllll}
  |w_1|^2 \left(1 + \frac{1}{4} K_4 |w_1|^2\right) &
  w_1^* \left(1 + \frac{1}{2} K_4 |w_1|^2\right) &
  \frac{1}{2} K_4 (w_1^*)^2 &
  \Re\left( w_1^* w_2 \left(1 + \frac{1}{4} K_4 w_1^* w_2\right)\right) &
  w_1^* \left(1 + \frac{1}{2} K_4 w_1^* w_2\right) &
  \frac{1}{2} K_4 (w_1^*)^2 \\
  &
  2 \left(1 + K_4 |w_1|^2\right) &
  2 K_4 w_1^* &
  w_2 \left(1 + \frac{1}{2} K_4 w_1^* w_2\right) &
  2 \left(1 + K_4 w_1^* w_2\right) &
  2 K_4 w_1^* \\
  &
  &
  2 K_4 &
  \frac{1}{2} K_4 w_2^2 &
  2 K_4 w_2 &
  2 K_4 \\
  &
  &
  &
  |w_2|^2 \left(1 + \frac{1}{4} K_4 |w_2|^2\right) &
  w_2^* \left(1 + \frac{1}{2} K_4 |w_2|^2\right) &
  \frac{1}{2} K_4 (w_2^*)^2 \\
  &
  &
  &
  &
  2 \left(1 + K_4 |w_2|^2\right) &
  2 K_4 w_2^* \\
  &
  &
  &
  &
  &
  2 K_4
\end{array}\right]
$$
Here $\Re$ denotes the real part.  The lower triangle can be filled in using the fact that the covariance matrix is Hermitian.  As before, there are correction terms of order $(\Rein/R_0)^4$ and higher.

There are two general conclusions to draw from this analysis.  First, the clump mass function enters the covariance matrix only through the effective mass, $\meff=\avg{m^2}/\avg{m}$.  Second, we see explicitly how the spatial distribution of clumps influences the long-range effects.  In covariances that only involve potential and/or deflection, the lowest-order term depends on
\begin{equation}
  K_2 = \int_{R_0}^{\infty} \frac{\ksub(r_i)}{r_i^2}\ r_i\ dr_i \\
\end{equation}
so distant clumps are effectively weighted by $1/r^2$.  Of course, this weighting is compensated to some extent by the fact that there is more area at large radius ($r\,dr$).  The net effect is that $K_2$ diverges logarithmically at large radius if the substructure population is spatially uniform, while $K_2 \propto R_0^{\eta-2}$ for the more realistic case of an asymptotic power law $\ksub \propto r^{\eta-2}$.  In covariances that involve a shear, the lowest-order term depends on
\begin{equation}
  K_2 K_4 = \int_{R_0}^{\infty} \frac{\ksub(r_i)}{r_i^4}\ r_i\ dr_i
\end{equation}
(The factor of $K_2$ comes from the leading factor in eq.~\ref{eq:covmat}.)  These terms clearly have less sensitivity to distant clumps.

\begin{table}
\caption{Images positions for the sample lens used in \reffig{longtest}.  The positions are in arcseconds relative to the centre of the lens galaxy, and the images are listed in time delay order.  See \citet{NewChannel} for more details.}
\begin{tabular}{crr}
\hline
 image & \multicolumn{1}{c}{$x$} & \multicolumn{1}{c}{$y$} \\
\hline
 M1 & $ 0.343$ & $ 1.360$ \\
 M2 & $-0.948$ & $-0.697$ \\
 S1 & $-1.098$ & $-0.206$ \\
 S2 & $ 0.700$ & $-0.652$ \\
\hline
\label{tab:foldlens}
\end{tabular}
\end{table}

We now use our fiducial clump model from \refsec{sample} to construct a quantitative example.  Since we are dealing with two-point statistics we must specify the positions of the images.  For illustration we use the sample ``fold'' lens defined by \citet{NewChannel}, which is similar to the observed lens PG~1115+080.  The images (which are listed in \reftab{foldlens}) correspond to a lens comprising a singular isothermal sphere with external shear, which has an Einstein radius $\Rein=1.16$ arcsec.  We run $10^3$ Monte Carlo simulations of clumps at radii $3\Rein < r < 100\Rein$, using a power law spatial distribution with $\eta=1$ normalised so $\ksub=0.01$ at the Einstein radius.  We also compute the analytic covariance matrix from \refeq{covmat}.

\begin{figure}
\centerline{\includegraphics[width=17.0cm]{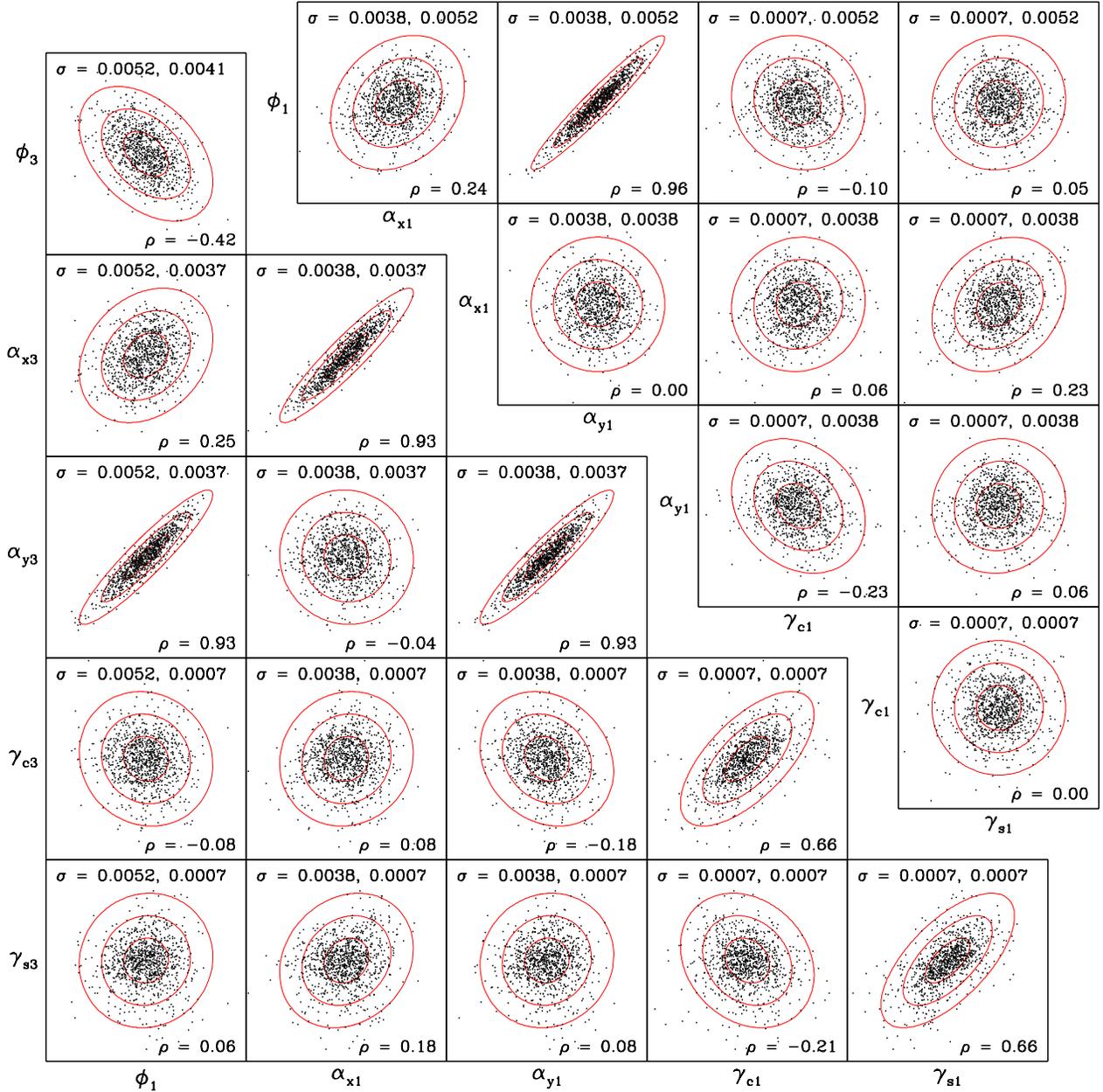}}
\caption{
Illustration of elements of the covariance matrix for long-range effects.  The upper triangle shows covariances among the potential, deflection, and shear for a single image (M1), while the lower triangle shows covariances between two different images (M1 and S1).  In each panel, the points show results from $10^3$ Monte Carlo simulations of clumps at $3\Rein < r < 100\Rein$, using our fiducial example in which the mass function has a dynamic range $q=100$ and the spatial distribution is a power law with $\eta=1$ and $\ksub=0.01$ at the Einstein radius.  The curves show the 1-, 2-, and 3-$\sigma$ contours predicted from the analytic covariance matrix (eq.~\ref{eq:covmat}).  The $\sigma$ values give the standard deviations in the horizontal (``$x$'') and vertical (``$y$'') directions, while $\rho$ gives the Pearson product-moment correlation coefficient for the two quantities: $\rho = \cov(x,y)/\sigma_x \sigma_y$.  The specific values of the correlation coefficients depend on the choice of inner radius ($R_0 = 3\Rein$ here).
}\label{fig:longtest}
\end{figure}

\reffig{longtest} shows covariances among the potential, deflection, and shear for a single image (upper triangle), as well as covariances between two different images (lower triangle).  For clarity we use the two real components of deflection and shear (instead of complex variables) in the example.  We have chosen images M1 and S1 for illustration, but the results are similar for other pairs.  One key result is that the direct simulations validate the analytic covariance matrix.  A second result is that there are non-trivial correlations among the lensing quantities.  For a single image, the degree to which the potential is correlated with the different components of the deflection depends on the position of the image; because of its position near the $y$-axis, image M1 has a stronger correlation between $\phi$ and $\alpha_y$ than between $\phi$ and $\alpha_x$.  For different images, the various correlations emerge because a distant clump can affect nearby images in a similar way; although with a large number of clumps there is always some stochasticity and the correlations are never perfect.  The correlations depend on the inner radius of the region containing clumps (the cut-off radius $R_0$ above): if we make $R_0$ large and consider only distant clumps, the correlations will be strong because (again) distant clumps affect the images in similar ways; whereas if we make $R_0$ small and consider relatively nearby clumps as well, the correlations will be weaker because nearby clumps affect the images differently.  It would be interesting in follow-up work to consider what physical information about the clump population is contained in the correlations among lensing quantities.  For now, we consider that this example has provided a useful demonstration of the concepts derived from our formal analysis of long-range effects.

%=====================================================================
\section{Discussion}
\label{sec:discussion}
%=====================================================================

\begin{figure}
\centerline{\includegraphics[width=17.0cm]{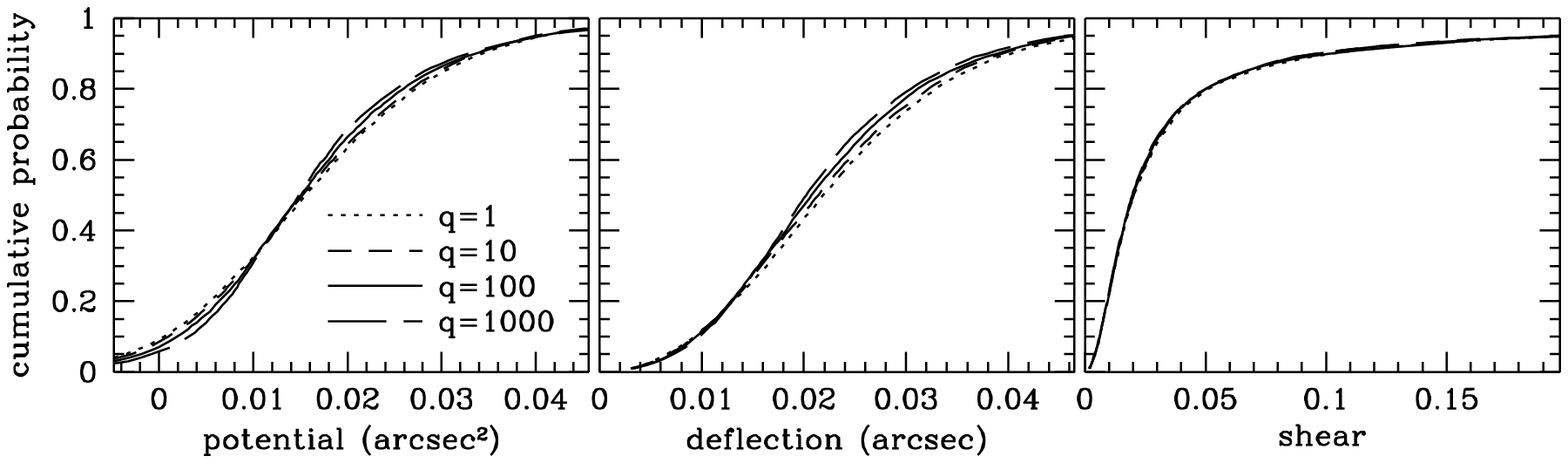}}
\caption{
Cumulative probability distributions for the total potential, deflection amplitude, and shear amplitude.  The different curves correspond to different mass functions, all normalised to have the same effective mass, $\meff=\avg{m^2}/\avg{m}$ (see \reftab{massfunc}).  The spatial distribution is a power law with $\eta=1$ and $\ksub=0.01$ at the position of the image.  We now plot the differential potential (relative to the origin), so the zeropoint is irrelevant.
}\label{fig:meff-tot}
\end{figure}

In our formal analysis we have so far considered clumps near an image (\refsec{local}) and clumps far away (\refsec{long}).  While we do not yet have a full theory for the total effects from all clumps (including those at intermediate distances),\footnote{Recall that the net effects from all clumps have been studied for the case of a uniform spatial distribution, most recently in the work by \citet{alberto1,alberto2}.} we can examine such effects numerically with our Monte Carlo simulations. Our goal in this section is to see whether the inferences we have drawn from the local and long-rage analyses extend to the effects from the full clump population.

One set of inferences involve the mass function of clumps, so we first study how different mass functions affect the probability distributions for the total potential, deflection, and shear.  \reffig{meff-tot} shows results from simulations with different mass functions when they are normalised to have the same effective mass, $\meff$.  (This is analogous to \reffig{meff-loc}, but now showing total substructure effects instead of just local effects.)  First consider the shear.  We have found that local effects dominate the shear, and local effects are insensitive to the clump mass function.  To the extent that long-range effects contribute to the shear, they introduce a dependence on $\meff$.  Thus, it is no surprise to see in \reffig{meff-tot} that the total shear distributions are basically indistinguishable for the different mass functions.  For comparison, previous analytic results for a uniform spatial distribution of point masses (i.e., microlensing) found that the shear distribution is strictly insensitive to the mass function, at least when the number of clumps is sufficiently large \citep{schneider-shear,alberto2}.  \citet{rozo} argued that the shear and magnification distributions can depend on the clump mass function when the clumps are spatially extended, although simulations of substructure lensing indicate that any dependence on the mass function is not strong \citep[when the source is small; e.g.,][]{DK02,ShinEvans}.

Next consider the deflection.  We have found that local and long-range effects are comparable in importance \citep[also see][]{chen-pos}.  The deflection distribution depends on the mass function only through $\meff$ in the case of a uniform spatial distribution; see \refeqs{Ploc2a}{covmat}, and also \citet{katz} and \citet{neindorf}.  Other mass moments enter when the spatial distribution is not uniform (see eq.~\ref{eq:Ploc5a}), but \reffig{meff-tot} shows that the deflection distribution is still only mildly sensitive to the mass function.  Finally consider the potential.  We have found that local effects in the potential are negligible, and long-range effects depend on the mass function through $\meff$.  This explains why the potential distribution shows only small variations with the mass function.  Now, it is an oversimplification to say that $\meff$ is the \emph{only} important property of the mass function, especially when discussing the deflection and potential; there are some residual variations in the probability distributions, which presumably arise from intermediate-scale clumps not yet treated in our formal theory, that will need to be incorporated into any detailed, quantitative study of substructure lensing.  Nevertheless, as a conceptual rule of thumb it seems fair to say that the deflection and potential are mainly sensitive to the effective clump mass.

\begin{figure}
\centerline{\includegraphics[width=17.0cm]{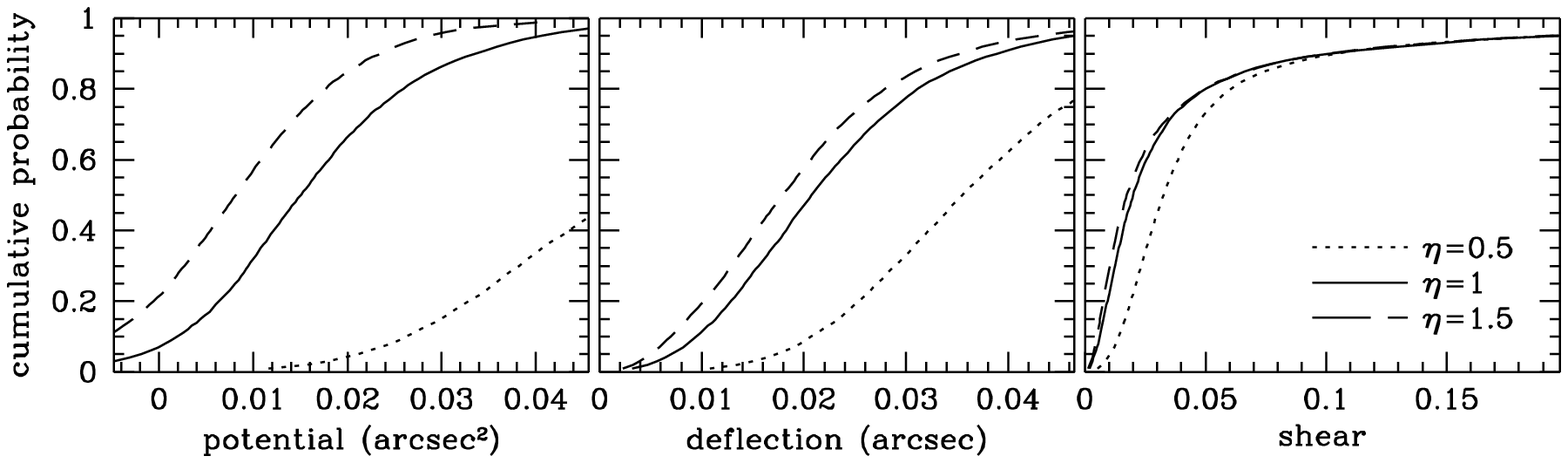}}
\caption{
Similar to \reffig{meff-tot}, but now for different power law indices for the spatial distribution of clumps: $\ksub \propto r^{\eta-2}$.  Results are shown for a mass function with dynamic range $q=100$.
}\label{fig:eta-tot}
\end{figure}

The other set of inferences involve the spatial distribution of clumps, so \reffig{eta-tot} shows probability distributions for the total potential, deflection, and shear from simulations with different power law radial profiles.  We have already seen that local effects are not very sensitive to the power law index $\eta$ (see \reffig{eta-loc}).  The total shear is still only modestly sensitive to $\eta$, and only at the low end of the shear distribution; the reason is that large shears are dominated by clumps in the vicinity of an image.  The total deflection is more sensitive to $\eta$, because it contains a significant contribution from non-local clumps.  And the total potential is very sensitive to the spatial distribution of clumps, since it is dominated by non-local effects.

%=====================================================================
\section{Conclusions}
\label{sec:conclusions}
%=====================================================================

We have developed certain aspects of the theory of gravitational lensing with stochastic substructure in order to understand how information about the population of mass clumps is encoded in various lensing observables.  Specifically, we have derived probability distributions for the potential, deflection, and shear produced by the clumps in the vicinity of a lensed image; and we have computed the covariance matrix for a multivariate Gaussian distribution representing the potential, deflection, and shear due to clumps far from the lensed images.  This analysis extends previous work on stochastic microlensing
\citep{nityananda,katz,deguchi1,deguchi2,schneider-shear,schneider-mu,seitz1,seitz2,neindorf,tuntsov1,tuntsov2,alberto1,alberto2} by allowing both general spatial distributions and mass functions for the clump population.

We have drawn two main conclusions about the clump mass function.  First, the probability distribution for the local shear is strictly independent of the clump mass function for a uniform spatial distribution, and essentially insensitive to the mass function for more general spatial distributions.  Since the total shear is dominated by clumps near the image, the probability distribution for the total shear is effectively insensitive to the mass function as well.  Second, the probability distributions for the potential and deflection depend on the mass function mainly through the effective mass, $\meff=\avg{m^2}/\avg{m}$.  There are some higher-order effects that depend on other moments of the mass function, but the principal scaling for both the potential and deflection is with $\meff$.  Our conclusions about the shear and deflection generalise previous results on stochastic microlensing with an infinite, uniform distribution of stars \citep{katz,schneider-shear,neindorf,alberto1,alberto2}.

The spatial distribution of clumps has little effect on the probability distributions for the local shear and potential, and only a modest effect on the probability distribution for the local deflection.  (It is no surprise, of course, that ``local'' effects are not very sensitive to the global distribution.)  Even the total shear distribution has only a modest sensitivity to the spatial distribution of clumps.  The total deflection distribution, by contrast, changes more significantly when we vary the spatial distribution; and the total potential distribution is very sensitive to changes in the global population of clumps.

The high-level conclusion from this work is that different lensing observables depend on the clump population in different ways, as summarised in \reftab{complementarity}.  If we can measure not only magnifications (in practice, magnification ratios) but also image positions and time delays well enough to detect substructure effects, we will gain the ability to probe substructure in lens galaxies in new ways and extract additional information about the clump populations.

\begin{table}
\caption{Heuristic guide to complementarity in substructure lensing.  In Column 2 we use the scaled radius $\rhat = r/\Rein$, where $r$ is the distance of a clump from the image and $\Rein$ is the Einstein radius of the clump (not of the macromodel).}
\begin{tabular}{|llcc|}
\hline
observable     & mnemonic                              & mass scale                   & spatial scale \\
\hline
magnifications & $\delta\gamma \sim 1/\rhat^2$         & $\int m\,\frac{dN}{dm}\,dm$  & quasi-local  \\
positions      & $\delta x     \sim \Rein/\rhat$       & $\avg{m^2}/\avg{m}$          & intermediate \\
time delays    & $\delta \phi  \sim \Rein^2 \ln \rhat$ & $\avg{m^2}/\avg{m}$          & long-range   \\
\hline
\label{tab:complementarity}
\end{tabular}
\end{table}

There are, to be sure, some limits to the analysis presented here.  First, as in previous work on stochastic microlensing, we have explicitly assumed point mass clumps and neglected any correlations between clumps.  While the point mass approximation is fine for clumps projected far from the lensed images, it will break down for clumps that overlap the line of sight; this will have the greatest effect on the substructure shear, since that is mainly associated with clumps near an image \citep[see][]{rozo}.  It will be interesting in future work to seek a general theory of stochastic lensing that can handle ``puffy'' clumps \citep[as in the substructure simulations by][]{MM01,DK02,maccio,chen-pos,ShinEvans} and allow correlations between clumps.  Second, this paper has implicitly focused on lensing of a point-like source.  There are ways to treat an extended source in the context of microlensing \citep[e.g.,][]{deguchi1,deguchi2,seitz1,seitz2,neindorf,tuntsov1,tuntsov2}, although the issues may be somewhat different in substructure lensing since the optical depth is generally lower and there can be complicated---and interesting---``resonances'' that appear when the source size is comparable to the Einstein radius of a clump \citep[see][]{dobler-finsrc}.  Once finite source effects are incorporated, this formalism could be applied to other problems in stochastic lensing such as planetesimal disk microlensing \citep{heng-disk1}.

There are several attractive opportunities to extend this work.  We have already noted that there can be correlations among the various lensing quantities (see \refsec{long}), and it would be interesting to see whether such two-point statistics contain additional information about the clump population.  Such an analysis could even be extended to higher-order statistics, although it is not clear when the data might be available to examine those.  Another goal would be to develop a more complete description of how the spatial distribution of clumps affects the various lensing quantities.  We can think of this in terms of a spatial kernel that gives different weights to clumps at different distances from an image.  We have identified the behavior of this kernel in the ``near'' and ``far'' regimes, but it would be nice to understand the full spatial kernel or at least identify the spatial moments that are most important for the total potential, deflection, and shear.

%=====================================================================
\appendix
\section{Averaging over the clump population}
\label{app:stats}
%=====================================================================

In this Appendix we specify how to compute averages over many realizations of the clump population.  Consider any quantity $f$ that is a sum of contributions from individual clumps, $f = \sum_i f_i$, where $i$ is the clump index.  The general average over clump realizations has the form:
\begin{equation} \label{eq:genavg}
  \avg{f} \equiv \int f\ \left\{\prod_j p_w(w_j)\ p_m(m_j)\ dw_j\ dm_j\right\}
  = \sum_i \int f_i\ p_w(w_i)\ p_m(m_i)\ dw_i\ dm_i
  = \frac{1}{\mbar} \int f_i\ \ksub(w_i)\ p_m(m_i)\ dw_i\ dm_i
\end{equation}
The first step is the formal definition of the average over clump populations.  In the second step we use the fact that $p_w(w_j)$ and $p_m(m_j)$ are normalised so they integrate to unity for $j \ne i$.  In the third step recognise that all terms in the sum are the same, so we can replace the sum with multiplication by $N$; and we use \refeq{p2kap} to rewrite $p_w(w) = \ksub(w)/N\avg{m}$.  Finally, since all the lensing quantities are proportional to the clump mass (cf.\ eqs.~\ref{eq:poti}--\ref{eq:shri}), we can define ${\hat f}_i = f_i/m_i$ to be a quantity that is independent of mass.  The factor of $m_i$ goes directly into the mass integration, which becomes $\int m_i\,p_m(m_i)\,dm_i = \avg{m}$, so we obtain
\begin{equation} \label{eq:avgf}
  \avg{f} = \int {\hat f}_i\ \ksub(w_i)\ dw_i
\end{equation}
Note that in this expression the $i$ is arbitrary; it simply indicates that when evaluating the remaining integral we  consider only an \emph{individual} clump.

Now we consider the covariance between two quantities $f = \sum_i f_i$ and $g = \sum_j g_j$.  We can write the covariance as follows (note the complex conjugation, which makes the covariance matrix Hermitian):
\begin{eqnarray}
  \cov(f,g) &\equiv& \avg{f g^*} - \avg{f}\avg{g^*} \\
  &\equiv& \left[\sum_{i=j} \int f_i\ g^*_i\ p_w(w_i)\ p_m(m_i)\ dw_i\ dm_i\right]
   + \left[\sum_{i \ne j} \int f_i\ g^*_j\ p_w(w_i)\ p_w(w_j)\ p_m(m_i)\ p_m(m_j)\ 
    dw_i\ dm_i\ dw_j\ dm_j\right] \nonumber\\
  && - \left[\sum_{i} \int f_i\ p_w(w_i)\ p_m(m_i)\ dw_i\ dm_i\right]
    \left[\sum_{j} \int g^*_j\ p_w(w_j)\ p_m(m_j)\ dw_j\ dm_j\right] \nonumber\\
  &=& N \left[\int f_i\ g^*_i\ p_w(w_i)\ p_m(m_i)\ dw_i\ dm_i\right]
   - N \left[\int f_i\ p_w(w_i)\ p_m(m_i)\ dw_i\ dm_i\right]
    \left[\int g^*_j\ p_w(w_j)\ p_m(m_j)\ dw_j\ dm_j\right] \nonumber\\
  &=& \frac{1}{\avg{m}} \int f_i\ g^*_i\ \ksub(w_i)\ p_m(m_i)\ dw_i\ dm_i
   - \frac{1}{N\avg{m}^2} \int f_i\ g^*_j\ \ksub(w_i)\ \ksub(w_j)\ p_m(m_i)\ p_m(m_j)\ 
    dw_i\ dm_i\ dw_j\ dm_j \nonumber
\end{eqnarray}
In the second step we write out the averages, using the fact that $p_w(w_k)$ and $p_m(m_k)$ integrate to unity for all $k \ne i,j$, and grouping terms in the double sum according to whether $i=j$ or $i \ne j$.  We then recognise that the second and third terms have identical forms, so we just need to determine how many total elements there are in these sums.  There are $N(N-1)$ elements in the second term (sum over $i \ne j$), and $N^2$ elements in the third term (full sums over both $i$ and $j$), yielding a total of $N$ elements with an overall minus sign.  Also, there are $N$ elements of the sum over $i=j$ in the first term.  This allows us to obtain the simplified expression in the third line.  Finally, we replace $p_x(w) = \ksub(w)/N\avg{m}$ to obtain the expression in the fourth line.  In this expression, note that the first term is independent of the number of clumps $N$, while the second term is $\order{1/N}{}$.  When the number of clumps is large, the second term is negligible and to good approximation we have
\begin{equation} \label{eq:covfg}
  \cov(f,g) \approx \frac{1}{\avg{m}} \int f_i\ g^*_i\ \ksub(w_i)\ p_m(m_i)\ dw_i\ dm_i
  = \frac{\avg{m^2}}{\avg{m}} \int {\hat f}_i\ {\hat g}^*_i\ \ksub(w_i)\ dw_i
\end{equation}
where (as before) we define ${\hat f}_i = f_i/m_i$ and ${\hat g}_i = g_i/m_i$ as quantities that are independent of mass.

%=====================================================================
\section*{Acknowledgments}
\label{S:acknowledgments}
%=====================================================================

I thank
Jacqueline Chen,
Art Congdon,
Greg Dobler,
Ross Fadely,
Mike Kuhlen,
Leonidas Moustakas,
Joel Primack,
Eduardo Rozo,
Kris Sigurdson, and
Risa Wechsler
for helpful discussions about astrophysical aspects of dark matter, substructure, and lensing; as well as
Arlie Petters,
Alberto Teguia, and
Kelly Wieand
for productive conversations about mathematical aspects of stochastic lensing.
Ross Fadely also provided helpful comments on the manuscript.
I thank the astrophysics group at the Institute for Advanced Study and the organisers and participants of the workshops ``Strong Gravitational Lensing in the Next Decade'' and ``Shedding Light on the Nature of Dark Matter'' for opportunities to present preliminary versions of this work.
Principal support for this work has come from NSF through grant AST-0747311, with additional support from the Institute for Advanced Study and the W.\ M.\ Keck Institute for Space Studies.

%=====================================================================
% REFERENCES
%=====================================================================

%\bibliographystyle{mn2e}
%\bibliography{stochastic}

\end{document}